\documentclass{ws-procs975x65}
\newcommand{\be}{\begin{equation}}
\newcommand{\ee}{\end{equation}}
\newcommand{\bea}{\begin{eqnarray}}
\newcommand{\eea}{\end{eqnarray}}
\newcommand{\nn}{\nonumber}
\newcommand{\bnabla}{\mbox{\boldmath{$\nabla$}}}
\begin{document}

\title{Local and Global Casimir Energies in a
Green's Function Approach}

\author{K. A. MILTON$^*$ and I. CAVERO-PEL\'AEZ$^\dagger$}

\address{Oklahoma Center for High Energy Physics and 
H. L. Dodge Department of Physics, University of Oklahoma,\\
Norman, OK 73019 USA\\
$^*$E-mail: milton@nhn.ou.edu\\
www.nhn.ou.edu/\%7Emilton\\
$^\dagger$E-mail: cavero@nhn.ou.edu}

\author{K. KIRSTEN}

\address{Department of Mathematics, Baylor University,\\
Waco, TX 76798 USA\\
E-mail: Klaus\_Kirsten@baylor.edu}

\begin{abstract}
The effects of quantum fluctuations in fields confined by background
configurations may be simply and transparently computed using the
Green's function approach pioneered by Schwinger.  Not only can
total energies and surface forces be computed in this way, but local
energy densities, and in general, all components of the vacuum
expectation value of the energy-momentum tensor may be calculated.
For simple geometries this approach may be carried out exactly, which
yields insight into what happens in less tractable situations.  In this
talk I will concentrate on the example of a scalar field in 
a circular cylindrical delta-function background. This situation is quite
similar to that of a spherical delta-function background. The local energy
density in these cases diverges as the surface of the background is 
approached, but these divergences are integrable.  The total energy is 
finite in strong coupling, but in weak coupling a divergence occurs
in third order. This universal feature is shown to reflect a divergence 
in the energy associated with the surface, the integrated local energy
density within the shell itself, which surface energy 
should be removable by a process of renormalization.
\end{abstract}


\bodymatter

\section{Casimir Energies for Spheres and Cylinders}\label{kam:sec1}
The calculation of Casimir self-energies of material objects has become
controversial,\cite{Graham:2003ib} although these concerns are nearly as
old as the subject itself.\cite{deutsch,candelas,miltonballs}  Although it
appears possible to extract unique self-energies, they may be overwhelmed
by terms which become divergent for ideal geometries.\cite{barton2k,barton03}
Our attitude is that these terms may be uniquely removed by a process
of renormalization, and that even the divergences revealed by heat-kernel
methods\cite{bkv,Gilkey:2001mj} may be unambiguously isolated.

Table \ref{kam:tab1} summarizes the state of our knowledge concerning total
Casimir self-energies for different simple configurations.  The first
row of the table refers to the Casimir energy of a perfectly conducting
shell, either spherical or cylindrical, subject to electromagnetic fluctuations
in the exterior and interior regions.  The second row refers to the same
results for a scalar field subject to Dirichlet boundary conditions on the
surface.  The remaining four rows describe small perturbations: Row 3 describes
what happens for electromagnetic fluctuations when the interior of the sphere 
or cylinder is a dielectric having a permittivity $\varepsilon$
differing slightly from the vacuum value of unity; Row 4 indicates the
same when the speed of light is the same inside and outside the object, where
$\xi=(\varepsilon'-\varepsilon)/(\varepsilon'+\varepsilon)$ in 
terms of the permittivity
inside ($\varepsilon$) and outside ($\varepsilon'$) the object; 
Row 5 shows the effect
for a perfect conductor of a small ellipticity $\delta e$
($\pm$ refers to a prolate or oblate spheroid, respectively); 
and Row 6 refers to a $\delta$-function
potential (semitransparent shell) of strength $\lambda$, which will be 
described in this paper.  In these four cases, what is shown in the table
is the coefficient of the second-order term in the relevant small quantity.
One of the ongoing challenges facing quantum field theorists attempting to
understand the quantum vacuum is to understand the pattern of signs and zeroes
manifested in the table.
\begin{table}
\tbl{Casimir energy ($E$) for a sphere and Casimir energy per unit length ($
\mathcal{E}$) for a cylinder, both of radius $a$. The signs indicate
repulsion or attraction, respectively.}
{\begin{tabular}{@{}cccc@{}}
\toprule
Type&$E_{\rm Sphere}a$&$\mathcal{E}_{\rm Cylinder}a^2$&References\\
\colrule
EM&$+0.04618$&$-0.01356$&\refcite{boyersphere}, \refcite{deraadcyl}\\
D&$+0.002817$&$+0.0006148$&\refcite{benmil}, \refcite{gos}\\
$(\varepsilon-1)^2$&$+0.004767=\frac{23}{1536\pi}$&$0$
&\refcite{bmm}, \refcite{Cavero-Pelaez:2004xp}\\
$\xi^2$&$+0.04974=\frac5{32\pi}$&$0$&\refcite{klich}, \refcite{dicyl}\\
$\delta e^2$&$\pm0.0009$&$0$&\refcite{Kitson:2005kk}, \refcite{Kitson:2006hf}\\
$\lambda^2$&$+0.009947=\frac1{32\pi}$&$0$
&\refcite{Milton:2002vm}, \refcite{Cavero-Pelaez:2006rt}\\ \botrule
\end{tabular}}
\label{kam:tab1}
\end{table}

In this talk, we will illustrate the ideas for the interesting case of
a circular cylindrically symmetric annular potential.  Most of the calculations
will refer to a $\delta$-function potential.

\section{Green's Function}
We consider a massless scalar field $\phi$ in a $\delta$-cylinder background,
\be
\mathcal{L}_{\rm int}=-\frac\lambda{2a}\delta(r-a)\phi^2,
\ee
$a$ being the radius of the ``semitransparent'' cylinder.  We recall that the
massive case was earlier considered by Scandurra.\cite{scancyl}  Note that
with this definition, $\lambda$ is dimensionless.
The time-Fourier transform of the Green's function,
\be
G(x,x')=\int\frac{d\omega}{2\pi}e^{-i\omega(t-t')}\mathcal{G}(\mathbf{r,r'}),
\ee
satisfies
\be
\left[-\nabla^2-\omega^2+\frac\lambda{a}\delta(r-a)\right]\mathcal{G}
(\mathbf{r,r'})=\delta(\mathbf{r-r'}).
\ee
Adopting cylindrical coordinates, we write
\be
\mathcal{G}(\mathbf{r,r'})=\int\frac{dk}{2\pi}e^{ik(z-z')}
\sum_{m=-\infty}^\infty
\frac1{2\pi}e^{im(\varphi-\varphi')}g_m(r,r';k),
\ee
where the reduced Green's function satisfies
\be
\left[-\frac1r\frac{d}{dr}r\frac{d}{dr}+\kappa^2+\frac{m^2}{r^2}
+\frac\lambda{a}
\delta(r-a)\right]g_m(r,r';k)=\frac1r\delta(r-r'),\label{kam:rgfeqn}
\ee
where $\kappa^2=k^2-\omega^2$.  Let us immediately make a Euclidean rotation,
\be
\omega\to i\zeta,
\ee
where $\zeta$ is real, so $\kappa$ is always real and positive.  Apart from
the $\delta$ functions, \eref{kam:rgfeqn} is the modified Bessel equation.

\subsection{Reduced Green's function}
Because of the Wronskian satisfied by the modified Bessel functions,
\be
K_m(x)I'_m(x)-K'_m(x)I_m(x)=\frac1x,\label{kam:wronskian}
\ee
we have the general solution to the Green's function equation
(\ref{kam:rgfeqn}) as long as $r\ne a$ to be
\be
g_m(r,r';k)=I_m(\kappa r_<)K_m(\kappa r_>)+A(r')I_m(\kappa r)+B(r')
K_m(\kappa r),
\ee
where $A$ and $B$ are arbitrary functions of $r'$. 
Now we incorporate the effect of the $\delta$ function at $r=a$
in the Green's function equation.
It implies that $g_m$ must be continuous at $r=a$, while it has a discontinuous
derivative,
\be
a\frac{d}{dr}g_m(r,r';k)\bigg|_{r=a-}^{r=a+}=\lambda g_m(a,r';k),
\ee
from which we rather immediately deduce the form of the Green's function
inside and outside the cylinder:
\begin{subequations}
\label{kam:g}
\bea
r,r'<a:\quad g_m(r,r';k)&=&I_m(\kappa r_<)K_m(\kappa r_>)\nonumber\\
&-&\frac{\lambda
K_m^2(\kappa a)}{1+\lambda I_m(\kappa a)K_m(\kappa a)}I_m(\kappa r)
I_m(\kappa r'),\label{kam:gin}
\\
r,r'>a:\quad g_m(r,r';k)&=&I_m(\kappa r_<)K_m(\kappa r_>)\nonumber\\
&-&\frac{\lambda
I_m^2(\kappa a)}{1+\lambda I_m(\kappa a)K_m(\kappa a)}K_m(\kappa r)
K_m(\kappa r').\label{kam:gout}
\eea
\end{subequations}
Notice that in the limit $\lambda\to\infty$ we recover the Dirichlet cylinder
result, that is, that $g_m$ vanishes at $r=a$.

\section{Pressure and Energy}
The easiest way to calculate the total energy is to compute the pressure on the
cylindrical walls due to the quantum fluctuations in the field.  This may be
computed, at the one-loop level, from the vacuum expectation value of the
stress tensor,
\be
\langle T^{\mu\nu}\rangle
=\left(\partial^\mu\partial^{\prime\nu}-\frac12 g^{\mu\nu}
\partial^\lambda\partial'_\lambda\right)\frac1i G(x,x')\bigg|_{x=x'}
-\xi
(\partial^\mu\partial^\nu-g^{\mu\nu}\partial^2)\frac1i G(x,x).\label{kam:st}
\ee
Here we have included the conformal parameter $\xi$, which is equal
to 1/6 for the conformal stress tensor. The conformal 
term does not contribute to the radial-radial component of the stress tensor, 
however, because then only transverse and time derivatives act on $G(x,x)$, 
which depends only on $r$.
  The discontinuity of the expectation value of the 
radial-radial   
component of the stress tensor is the pressure on the cylindrical wall:
\bea
P&=&\langle T_{rr}\rangle_{\rm in}-\langle T_{rr}\rangle_{\rm out}\nonumber\\
&=&-\frac1{16\pi^3}\sum_{m=-\infty}^\infty \int_{-\infty}^\infty dk
\int_{-\infty}
^\infty d\zeta\frac{\lambda \kappa^2}{1+\lambda I_m(\kappa a)K_m(\kappa a)}
\nonumber\\
&&\qquad\times
\left[K_m^2(\kappa a)I_m^{\prime 2}(\kappa a)-I_m^2(\kappa a)K_m^{\prime 2}
(\kappa a)\right]\nonumber\\
=&-&\frac1{16\pi^3}\sum_{m=-\infty}^\infty\int_{-\infty}^\infty\!\! dk
\int_{-\infty}
^\infty\!\! d\zeta\frac\kappa{a}\frac{d}{d\kappa a}\ln\left[1+\lambda I_m(\kappa a)
K_m(\kappa a)\right],
\eea
where we've again used the Wronskian (\ref{kam:wronskian}). 
 Regarding $ka$ and
$\zeta a$ as the two Cartesian components of a two-dimensional vector, with
magnitude $x\equiv\kappa a=\sqrt{k^2a^2+\zeta^2 a^2}$, we get the stress on the
cylinder per unit length to be
\be
\mathcal{S}=2\pi a P=-\frac1{4\pi a^3}\int_0^\infty
\!\! dx\,x^2\sum_{m=-\infty}^\infty
\frac{d}{dx}\ln\left[1+\lambda I_m(x)K_m(x) \right],
\ee
implying the Dirichlet limit as $\lambda\to\infty$.
By integrating
$\mathcal{S}=-\frac\partial{\partial a}\mathcal{E}$,
we obtain the energy per unit length
\be
\mathcal{E}=-\frac1{8\pi a^2}\int_0^\infty dx\,x^2\sum_{m=-\infty}^\infty
\frac{d}{dx}\ln\left[1+\lambda I_m(x)K_m(x) \right].\label{kam:energy}
\ee
This formal expression will be regulated, and evaluated in weak and strong
coupling, in the following.

\subsection{Energy}
\label{kam:sec:energy}
Alternatively, we may compute the energy directly from the general
formula\cite{Milton:2004vy}
\be
E=\frac1{2i}\int(d\mathbf{r})\int\frac{d\omega}{2\pi}2\omega^2\mathcal{G}
(\mathbf{r,r}).
\ee
To evaluate the energy in this case, we need the indefinite integrals
\begin{subequations}\label{kam:ints}
\bea
\int_0^x dy\,y \,I_m^2(y)&=&
\frac12\left[(x^2+m^2)I_m^2(x)-x^2 I_m^{\prime 2}\right],\label{kam:int1}
\\
\int_x^\infty dy\,y \,K_m^2(y)&=&-\frac12\left[(x^2+m^2)K_m^2(x)-
x^2 K_m^{\prime 2}\right].\label{kam:int2}
\eea
\end{subequations}
When we insert the above construction (\ref{kam:g}) 
of the Green's function, and perform
the integrals as indicated over the regions interior and exterior to the
cylinder, we obtain
\be
\mathcal{E}=-\frac{a^2}{8\pi^2}\sum_{m=-\infty}^\infty
\int_{-\infty}^\infty d\zeta\int_{-\infty}^\infty dk\,
\zeta^2\frac1x\frac{d}{dx}\ln\left[1+\lambda I_m(x)K_m(x)\right].
\ee
Again we regard the two integrals as over Cartesian coordinates, and replace
the integral measure by
\be
\int_{-\infty}^\infty d\zeta\int_{-\infty}^\infty dk \,\zeta^2=\pi\int_0^\infty
d\kappa\,\kappa^3.\label{kam:angint}
\ee
The result for the energy (\ref{kam:energy}) immediately follows. 

\section{Weak-coupling Evaluation}

Suppose we regard $\lambda$ as a small parameter, so let us expand 
the energy (\ref{kam:energy}) in powers of $\lambda$.  The first term is
\be
\mathcal{E}^{(1)}=-\frac\lambda{8\pi a^2}\sum_{m=-\infty}^\infty \int_0^\infty
dx\,x^2\,\frac{d}{dx}K_m(x)I_m(x).\label{kam:1storder}
\ee
The addition theorem for the modified Bessel functions is
\be
K_0(kP)=\sum_{m=-\infty}^\infty e^{im(\phi-\phi')}K_m(k\rho)I_m(k\rho'),\quad
\rho>\rho',\label{kam:addthm}
\ee
where $P=\sqrt{\rho^2+\rho^{\prime2}-2\rho\rho'\cos(\phi-\phi')}$. 
If this
is extrapolated to the limit $\rho'=\rho$ we conclude that the sum of the
Bessel functions appearing in $\mathcal{E}^{(1)}$ is $K_0(0)$, that is, a 
constant, so there is  no first-order contribution to the energy,
$\mathcal{E}^{(1)}=0$.

\subsection{Regulated numerical evaluation of $\mathcal{E}^{(1)}$}
\label{kam:secreg}
Given that the above argument evidently formally omits divergent terms, it may
be more satisfactory to offer a regulated numerical evaluation
of $\mathcal{E}^{(1)}$.  We can very efficiently do so using the uniform
asymptotic expansions ($m\to\infty$):
\begin{subequations}
\label{kam:uae}
\bea
I_m(x)&\sim&\sqrt{\frac{t}{2\pi m}}e^{m\eta}\left(1+\sum_{k=1}^\infty
\frac{u_k(t)}{m^k}\right),\label{kam:uaei}\\
K_m(x)&\sim&\sqrt{\frac{\pi t}{2 m}}e^{-m\eta}\left(1+\sum_{k=1}^\infty(-1)^k
\frac{u_k(t)}{m^k}\right),\label{kam:uaek}
\eea
\end{subequations}
where $x=mz$, $t=1/\sqrt{1+z^2}$, and $\frac{d\eta}{dz}=\frac1{zt}$.
The polynomials in $t$ appearing here are generated by
\bea
u_0(t)=1,\quad
u_k(t)=\frac12 t^2(1-t^2)u'_{k-1}(t)+\int_0^t ds\frac{1-5s^2}{8}u_{k-1}(s).
\eea
Thus the asymptotic behavior of the products of Bessel functions appearing
in \eref{kam:1storder} is obtained from
\be
I_m^2(x)K_m^2(x)\sim\frac{t^2}{4m^2}\left(1+\sum_{k=1}^\infty \frac{r_k(t)}
{m^{2k}}\right).\label{kam:asymprod}
\ee
The first three polynomials occurring here are
\begin{subequations}
\label{kam:rs}
\bea
r_1(t)&=&\frac{t^2}4(1-6t^2+5t^4),\\
r_2(t)&=&\frac{t^4}{16}(7-148t^2+554 t^4-708 t^6+295 t^8),\\
r_3(t)&=&\frac{t^6}{16}(36-1666t^2+13775t^4-44272t^6\nonumber\\
&&\quad\mbox{}+67162t^8-48510t^{10}
+13475 t^{12}).
\eea
\end{subequations}

We regulate the sum and integral by inserting an exponential cutoff, $\delta
\to 0+$:
\be
\mathcal{E}^{(1)}=-\frac{\lambda}{4\pi a^2}\sum_{m=0}^\infty{}'
\int_0^\infty dx\,x^2\frac{d}{dx}I_m(x)K_m(x) e^{-x\delta},
\ee
where the prime on the summation sign means that the $m=0$ term is counted
with one-half weight. We break up this expression into five parts,
\be
\mathcal{E}^{(1)}=-\frac{\lambda}{8\pi a^2}(\mbox{I}+\mbox{II}+\mbox{III}+
\mbox{IV}+\mbox{V}).
\ee
The first term is the $m=0$ contribution, suitably subtracted to make it
convergent (so the convergence factor may be omitted),
\be
\mbox{I}=\int_0^\infty dx\,x^2\frac{d}{dx}\left[I_0(x)K_0(x)
-\frac1{2\sqrt{1+x^2}}\right]=-1.\label{kam:i}
\ee
The second term is the above subtraction,
\be
\mbox{II}=\frac12\int_0^\infty dx\,x^2\left(\frac{d}{dx}\frac1{\sqrt{1+x^2}}
\right)e^{-x\delta}\sim-\frac1{2\delta}+1,\label{kam:ii}
\ee
as may be verified by breaking the integral in two parts at $\Lambda$,
$1\ll\Lambda\ll1/\delta$.
The third term is the sum over the $m$th Bessel function with the two
leading asymptotic approximants in \eref{kam:asymprod} subtracted: 
\be
\mbox{III}=
2\sum_{m=1}^\infty \int_0^\infty dx\,x^2\frac{d}{dx}\bigg[I_m(x)K_m(x)
-\frac{t}{2m}\left(1+\frac{t^2}{8m^2}(1-6t^2+5t^4)\right)\bigg]=0.
\label{kam:iii}
\ee
Numerically, each term in the sum seems to be zero to machine accuracy.
This is verified by computing the higher-order terms in that expansion,
in terms of the polynomials in Eq.~(\ref{kam:rs}):
\bea
&&I_m(x)K_m(x)
-\frac{t}{2m}\left(1+\frac{t^2}{8m^2}(1-6t^2+5t^4)\right)\nn\\
&\sim&
\frac{t}{4m^5}\left[r_2(t)-\frac14 r_1^2(t)\right]+\frac{t}{4m^7}\left[r_3(t)
-\frac12 r_1(t)r_2(t)+\frac18r_1^3(t)\right]+\dots,
\eea
which terms are easily seen to integrate to zero.
The fourth term is the leading subtraction which appeared in the third term:
\be
\mbox{IV}=\sum_{m=1}^\infty m\int_0^\infty dz\,z^2
\left(\frac{d}{dz}t\right) e^{-mz\delta}.
\ee
If we first carry out the sum on $m$ we obtain
\be
\mbox{IV}=
-\frac14\int_0^\infty dz\,z^3\frac1{(1+z^2)^{3/2}}\frac1{\sinh^2z\delta/2}
\sim-\frac1{\delta^2}+\frac1{2\delta}-\frac16,\label{kam:iv}
\ee
as verified by breaking up the integral.
The final term, due to the subleading subtraction,
if unregulated, is the form of infinity times zero:
\be
\mbox{V}=
\frac18\sum_{m=1}^\infty\frac1m\int_0^\infty dz\,z^2\frac{d}{dz}(t^3-6t^5+
5t^7)e^{-m z\delta}.
\ee
Here the sum on $m$ gives
\be
\sum_{m=1}^\infty \frac1m e^{-mz\delta}=-\ln\left(1-e^{-z\delta}\right),
\ee
and so we can write
\be
\mbox{V}=
\frac1{16}\frac{d}{d\alpha}\int_0^1 du\,(1-u)^\alpha u^{-2-\alpha}(u^{3/2}
-6u^{5/2}+5u^{7/2})\bigg|_{\alpha=0}=\frac16.\label{kam:v}
\ee
Adding together these five terms we obtain
\be
\mathcal{E}^{(1)}=\frac{\lambda}{8\pi a^2\delta^2}+0,
\ee
that is, the $1/\delta$ and constant terms cancel.
The remaining divergence
may be interpreted as an irrelevant constant, since $\delta=\tau/a$, $\tau$
being regarded as a point-splitting parameter.  This thus agrees with the
result stated at the beginning of this section.

\subsection{$\lambda^2$ term}
We can proceed the same way to evaluate the second-order contribution
to \eref{kam:energy},
\be
\mathcal{E}^{(2)}=\frac{\lambda^2}{16\pi a^2}\int_0^\infty dx\,x^2\,
\frac{d}{dx}\sum_{m=-\infty}^\infty I_m^2(x)K_m^2(x).\label{kam:2ndorden}
\ee
By squaring the sum rule (\ref{kam:addthm}), and again taking the formal singular limit
$\rho'\to\rho$, we evaluate the sum over Bessel functions appearing here
as
\be
\sum_{m=-\infty}^\infty I_m^2(x)K_m^2(x)=\int_0^{2\pi}\frac{d\varphi}{2\pi}
K_0^2(2x\sin\varphi/2).
\ee
Then changing the order of integration, we can write the second-order energy 
as
\be
\mathcal{E}^{(2)}=-\frac{\lambda^2}{64\pi^2 a^2}\int_0^{2\pi}\frac{d\varphi}
{\sin^2\varphi/2}\int_0^\infty dz\,z\,K_0^2(z),\label{kam:phive2}
\ee
where the Bessel-function integral has the value 1/2.  However, the integral
over $\varphi$ is divergent.  We interpret this integral by adopting
 an analytic regularization based on the integral ($\mbox{Re}\,s>-1$)
\be
\int_0^{2\pi}d\varphi \left(\sin\frac\varphi2\right)^s=\frac{2\sqrt{\pi}\Gamma
\left(\frac{1+s}2\right)}{\Gamma\left(1+\frac{s}2\right)}.
\ee
Taking the right-side of this equation
to define the $\varphi$ integral for all $s$, we conclude that the 
$\varphi$ integral, 
and hence the second-order energy $\mathcal{E}^{(2)}$, is zero.

The vanishing of the energy in order $\lambda$ and $\lambda^2$
may be given a quite rigorous derivation in the zeta-function approach
to Casimir energies---See Ref.~\refcite{Cavero-Pelaez:2006rt}.

\subsubsection{Alternative numerical evaluation}
Again we provide a numerical approach to bolster our argument.  Subtracting
and adding the leading asymptotic behaviors, 
we now write the second-order energy as ($z=x/m$)
\bea
\mathcal{E}^{(2)}&=&-\frac{\lambda^2}{8\pi a^2}\Bigg\{\int_0^\infty
dx\,x\left[I_0^2(x)K_0^2(x)-\frac1{4(1+x^2)}\right]\nonumber\\
&&\quad\mbox{}+\frac12\lim_{s\to 0}\sum_{m=0}^\infty{}'m^{-s}\int_0^\infty dz
\frac{z^{1-s}}{1+z^2}+
2\int_0^2 dz\,z\frac{t^2}4\sum_{m=1}^\infty \sum_{k=1}^3 \frac{r_k(t)}
{m^{2k}}\nonumber\\
&&\mbox{}+2\sum_{m=1}^\infty
\int_0^\infty dx\,x\left[I_m^2(x)K_m^2(x)-\frac{t^2}{4m^2}
\left(1+\sum_{k=1}^3\frac{r_l(t)}{m^{2k}}\right)\right]\Bigg\}.\label{kam:num}
\eea
The successive terms are evaluated as
\bea
\mathcal{E}^{(2)}&\approx&-\frac{\lambda^2}{8\pi a^2}\Bigg[
\frac14(\gamma+\ln4)-\frac14\ln2\pi-\frac{\zeta(2)}{48}
+\frac{7\zeta(4)}{1920}-\frac{31\zeta(6)}{16128}\nonumber\\
&&\quad+0.000864+0.000006\Bigg]=-\frac{\lambda^2}{8\pi a^2}(0.000000),
\eea
where  in the last term in the energy (\ref{kam:num}) 
only the $m=1$ and 2 terms are 
significant. Therefore, we have demonstrated numerically
that the energy in order $\lambda^2$ is zero to an accuracy of better than 
$10^{-6}$.
\subsubsection{Exponential regulator}
The astute listener will note that we used a standard, but possibly
questionable, analytic regularization in defining the second term in
energy above.  Alternatively, as in \sref{kam:secreg} we could insert
there an exponential regulator in each integral of $e^{-x\delta}$, with
$\delta$ to be taken to zero at the end of the calculation.  For $m\ne0$
$x$ becomes $mz$, and then the sum on $m$ becomes
\be
\sum_{m=1}^\infty e^{-mz\delta}=\frac1{e^{z\delta}-1}.
\ee
Then when we carry out the integral over $z$ we obtain for that term
\be
\frac\pi{8\delta}-\frac14\ln2\pi.\label{kam:cutoff1}
\ee
Thus we obtain the same finite part as above, but in addition an explicitly
divergent term
\be
\mathcal{E}^{(2)}_{\rm div}=-\frac{\lambda^2}{64 a^2\delta}.
\ee
Again, if we think of the cutoff in terms of a vanishing proper time $\tau$,
$\delta=\tau/a$, this divergent term is proportional to $1/a$, so the
divergence in the energy goes like $L/a$, if $L$ is the (very large) length
of the cylinder.  This is of the form of the {\em shape divergence} 
encountered in Ref.~\refcite{Cavero-Pelaez:2004xp}.

\subsection{Divergence in $O(\lambda^3)$}

Although the first two orders in $\lambda$ identically vanish, a
divergence in the energy (\ref{kam:energy}) does occur in $O(\lambda^3)$.  
\bea
\mathcal{E}^{(3)}&=&-\frac1{8\pi a^2}\sum_{m=-\infty}^\infty \int_0^\infty
dx\,x^{2-s}\frac{d}{dx}\frac13\lambda^3K_m^3(x)I_m^3(x)\nn\\
&\sim&\frac{\lambda^3}{96\pi a^2s},\quad s\to 0.\label{kam:thirddiv}
\eea
That such a divergence does occur generically in third order was proved
in Ref.~\refcite{Cavero-Pelaez:2006rt}, using heat-kernel techniques.
As we shall see, this divergence entirely arises from the 
{\em surface energy}.

\section{Strong Coupling}

The strong-coupling limit of the energy (\ref{kam:energy}), that is, the
Casimir energy of a Dirichlet cylinder, 
\be
\mathcal{E}^D=-\frac1{8\pi a^2}\sum_{m=-\infty}^\infty \int_0^\infty dx\,x^2
\frac{d}{dx}\ln I_m(x)K_m(x),
\ee
was worked out to high accuracy by Gosdzinski and Romeo,\cite{gos}
\be
\mathcal{E}^D=\frac{0.000614794033}{a^2}.\label{kam:grresult}
\ee
It was later redone with less accuracy by Nesterenko and 
Pirozhenko.\cite{nestcyl}
For completeness, let us sketch the evaluation here.  Again subtracting
and adding the leading asymptotics, we find for the energy per unit length
\bea
\mathcal{E}^D&=&-\frac1{8\pi a^2}\Bigg\{-2\int_0^\infty \!\!\!\!dx\,x\left[
\ln\left(2xI_0(x)K_0(x)\right)-\frac18\frac1{1+x^2}\right]\nonumber\\
&&\mbox{}+2\sum_{m=1}^\infty\int_0^\infty \!\!\!\!dx\,x^2\frac{d}{dx}
\left[\ln \left(2x I_m(x)K_m(x)
\right)-\ln\left(\frac{xt}{m}\right)-\frac12\frac{r_1(t)}{m^2}\right]
\nonumber\\
&&\mbox{}-2\sum_{m=0}^\infty{}'\int_0^\infty \!\!\!\!dx\,x^2\frac{d}{dx}\ln 2x
+2\sum_{m=1}^\infty \int_0^\infty \!\!\!\!dx\,x^2 \frac{d}{dx}\ln xt
\nonumber\\
&&+\sum_{m=1}^\infty \int_0^\infty \!\!\!\!dx\,x^2 \frac{d}{dx}\left[
\frac{r_1(t)}{m^2}-\frac14\frac1{1+x^2}\right]
-\frac12\sum_{m=0}^\infty{}'\int_0^\infty \!\!\!\!
dx\frac{x}{1+x^2}\Bigg\}.
\label{kam:scints}\eea
In the first two terms we have subtracted the leading asymptotic behavior so
the resulting integrals are convergent.  Those terms are restored in the 
fourth, fifth, and sixth terms.  The most divergent part of the Bessel 
functions
are removed by the insertion of $2x$ in the corresponding integral, and
its removal in the third term.  (Elsewhere, such terms have been referred to
as ``contact terms.'')  
The terms involving Bessel functions are 
evaluated numerically, where it is observed that the asymptotic value of the
summand (for large $m$)
in the second term is $1/32m^2$.  The fourth term is evaluated by writing it
as
\be
2\lim_{s\to0}\sum_{m=1}^\infty m^{2-s}\int_0^\infty dz
\frac{z^{1-s}}{1+z^2}=2\zeta'(-2)=-\frac{\zeta(3)}{2\pi^2},\label{kam:4th}
\ee
while the same argument, as anticipated, shows that the third ``contact'' term
is zero.\footnote{This argument is a bit suspect, since the
analytic continuation that defines the integrals has no common region
of existence.  Thus the argument in the following subsection may be
preferable.}
The sixth term is
\be
-\frac12\lim_{s\to 0}\left[\zeta(s)+\frac12\right]\frac1s=\frac14\ln 2\pi.
\ee
The fifth term is elementary.  The result then is
\bea
\mathcal{E}^D&=&\frac1{4\pi a^2}(0.010963-0.0227032+0+0.0304485+0.21875
-0.229735)\nn\\
&=&\frac{0.0006146}{a^2},\label{kam:myresult}
\eea
which agrees with \eref{kam:grresult} to the fourth significant 
figure.

\subsection{Exponential regulator}
As in the weak-coupling calculation, it may seem more satisfactory to insert an
exponential regulator rather than use analytic regularization.  Now it is
the third, fourth, and sixth terms in the above expression that must be 
treated. The latter is just the negative of the term 
(\ref{kam:cutoff1}) encountered in weak
coupling.  We can combine the third and fourth terms to give
\be
-\frac1{\delta^2}+\frac2{\delta^2}\int_0^\infty \frac{dz\,z^3}{z^2+\delta^2}
\frac{d^2}{dz^2}\frac{1}{e^z-1}.
\ee
The latter integral may be evaluated by writing it as an integral along
the entire $z$ axis, and closing the contour in the upper half plane,
thereby encircling the poles at $i\delta$ and at $2in\pi$, where $n$ is a
positive integer.  The residue theorem then gives for that integral
\be
-\frac{2\pi}{\delta^3}-\frac{\zeta(3)}{2\pi^2},
\ee
so once again, comparing with \eref{kam:4th}, 
we obtain the same finite part as  in \eref{kam:myresult}.

In this way of proceeding, then, in addition to the finite part 
found before in \eref{kam:myresult}, we obtain divergent terms
\be
\mathcal{E}^D_{\rm div}=\frac1{64a^2\delta}+\frac1{8\pi a^2\delta^2}
+\frac1{4a^2\delta^3},
\ee
which, with the previous interpretation for $\delta$, implies
terms in the energy proportional to $L/a$  (shape), $L$ (length),
and $aL$ (area), respectively, and are therefore {\em renormalizable}.
Had a logarithmic
divergence occurred (as does occur in weak coupling in $O(\lambda^3)$) such a
renormalization would be impossible.  However, see below!

\section{Local Energy Density}
We compute the energy density from the stress tensor (\ref{kam:st}), or
\bea
\langle T^{00}\rangle&=&\frac1{2i}\left(\partial^0\partial^{0\prime}
+\bnabla\cdot\bnabla'\right)G(x,x')\bigg|_{x'=x}-\frac\xi{i}\nabla^2G(x,x)
\nonumber\\
&=&\frac1{16\pi^3i}\int_{-\infty}^\infty dk\int_{-\infty}^\infty d\omega
\sum_{m=-\infty}^\infty \Bigg[\left(\omega^2+k^2+\frac{m^2}{r^2}
+\partial_r\partial_{r'}\right)g(r,r')\bigg|_{r'=r}\nn\\
&&\qquad\mbox{}-2\xi\frac1r\partial_r r\partial_r g(r,r)\Bigg].
\eea
We omit the free part of the Green's function (\ref{kam:g}), 
since that corresponds to the vacuum energy in the absence of the cylinder. 
When we insert the remainder of the Green's function, we obtain
the following expression for the energy density outside the cylindrical
shell:
\bea
u(r)&=&-\frac\lambda{16\pi^3}
\int_{-\infty}^\infty
d\zeta\int_{-\infty}^\infty dk\sum_{m=-\infty}^\infty \frac{
I_m^2(\kappa a)}{1+\lambda I_m(\kappa a)K_m(\kappa a)}
\nonumber\\
&&\times\Bigg[\left(2\omega^2+\kappa^2+\frac{m^2}{r^2}\right)K_m^2(\kappa
r)+\kappa^2K_m^{\prime 2}(\kappa r)\nonumber\\ 
&&\quad\mbox{}-2\xi\frac1r\frac\partial{\partial r}
r\frac\partial{\partial r} K_m^2(\kappa r)\Bigg],\qquad r>a.\label{kam:uofr}
\eea
The factor in square brackets can be easily seen to be, from the
modified Bessel equation, 
\be
2\omega^2 K_m^2(\kappa r)+\frac{1-4\xi}2\frac1r\frac\partial{\partial r}
r\frac\partial{\partial r} K_m^2(\kappa r).\label{kam:divform}
\ee
For the interior region, $r<a$, we have the corresponding expression for
the energy density with $I_m\leftrightarrow K_m$.

\subsection{Total and surface energy}
We first need to verify that we recover the expression for the energy
found before.  So let us integrate the above expression
over the region exterior of the cylinder, and the corresponding
interior expression over the inside region.
The second term in \eref{kam:divform}
is a total derivative, while the first may be integrated
according to the integrals given in \eref{kam:ints}.  
In fact that term is exactly that evaluated above.  The result is
\bea
\int\!\!(d\mathbf{r})\,u(r)&=&-\frac1{8\pi a^2}\sum_{m=-\infty}^\infty
\int_0^\infty\!\! dx\,x^2\frac{d}{dx}\ln\left[1+\lambda I_m(x)K_m(x)\right]
\nn\\
&&\mbox{}-(1-4\xi)\frac\lambda{4\pi a^2}
\int_0^\infty\!\!\!\!dx\,x 
\sum_{m=-\infty}^\infty \frac{I_m(x)K_m(x)}{1+\lambda I_m(x)K_m(x)}.
\label{kam:inten}
\eea
The first term is the total energy (\ref{kam:energy}), but what do we make of
the second term?  In strong coupling, it would represent a constant that
should have no physical significance (a contact term---it is independent
of $a$ if we revert to the physical variable $\kappa$ as the integration
variable).  

In general, however,
there is another contribution to the total energy, residing precisely
on the singular surface.  This surface energy is given in general
by\cite{dowkerandkennedy,kcd,Romeo:2000wt,Romeo:2001dd,Milton:2004vy,
Fulling:2003zx} 
\be
\mathfrak{E}=-\frac{1-4\xi}{2i}\oint_S d\mathbf{S}\cdot\bnabla G(x,x')
\bigg|_{x'=x},\label{kam:surfenergy}
\ee
which turns out to be
the negative of the second term in $\int(d\mathbf{r})\,u(r)$ given in 
Eq.~(\ref{kam:inten}).
This is an example of the general theorem
\be
\int(d\mathbf{r}) \,u(\mathbf{r})+\mathfrak{E}=E,
\ee
that is, the total energy $E$
is the sum of the integrated local energy density
and the surface energy.  A consequence of this theorem is that the
 total energy, unlike the local energy density, is independent of
the conformal parameter $\xi$.

\subsection{Surface divergences}
We now turn to an examination of the behavior of the local energy
density as $r$ approaches $a$ from outside the cylinder.
To do this we use the uniform asymptotic expansion (\ref{kam:uae}).  
Let us begin by
considering the strong-coupling limit, a Dirichlet cylinder.  If we stop with 
only the leading asymptotic behavior, we obtain the expression ($z=\kappa r/m$)
\be
u(r)\sim-\frac1{8\pi^3}\int_0^\infty d\kappa\,\kappa\sum_{m=-\infty}^\infty
e^{-m\chi}\left[-\kappa^2\frac{\pi t}{2m}
+2(1-4\xi)\kappa^2\frac{\pi}{2mt}\frac1{z^2}
\right],\qquad (\lambda\to\infty)
\ee
where
\be
\chi=-2\left[\eta(z)-\eta\left(z\frac{a}r\right)\right],
\ee
and we have carried out the ``angular'' integral as in \eref{kam:angint}.
Here we ignore the difference between $r$ and $a$ except in the exponent, and
we now replace $\kappa$ by $m z/a$.  
Close to the surface,
\be
\chi\sim \frac2t\frac{r-a}r,
\ee
and we carry out the sum over $m$ according to
\be
2\sum_{m=1}^\infty m^3 e^{-m\chi}\sim-2\frac{d^3}{d\chi^3}\frac1\chi
=\frac{12}{\chi^4}\sim\frac34\frac{t^4r^4}{(r-a)^4}.
\ee
Then the energy density behaves, as $r\to a+$, 
\bea
u(r)&\sim&-\frac1{16\pi^2}\frac1{(r-a)^4}(1-6\xi).
\eea
This is the universal surface divergence first discovered by Deutsch
and Candelas.\cite{deutsch}  It therefore occurs, with precisely the
same numerical coefficient, near a Dirichlet plate\cite{Milton:2002vm} 
or a Dirichlet sphere.\cite{Cavero-Pelaez:2005kq}  
It is utterly without physical significance (in the absence of gravity), 
and may be eliminated with the conformal choice for the
parameter $\xi$, $\xi=1/6$.

\subsection{Conformal surface divergence}\label{kam:secconf}
We will henceforth make this conformal choice.  Then the leading divergence
depends upon the curvature.  This was also worked out by Deutsch and
Candelas;\cite{deutsch} for the case of a cylinder, that result is
\be
u(r)\sim \frac1{720\pi^2}\frac1{r(r-a)^3},\quad r\to a+,\label{kam:dccyl}
\ee
exactly 1/2 that for a Dirichlet sphere of radius $a$.
To get this result, we keep the
$1/m$ corrections in the uniform asymptotic expansion, 
and the next term in $\chi$:
\be
\chi\sim\frac2t\frac{r-a}r+t\left(\frac{r-a}r\right)^2.\label{kam:chiexp}
\ee
\subsection{Weak coupling}

Let us now expand the energy density (\ref{kam:uofr}) for small coupling,
\bea
u(r)&=&-\frac{\lambda}{16\pi^3}\int_{-\infty}^\infty d\zeta
\int_{-\infty}^\infty dk\sum_{m=-\infty}^\infty I_m^2(\kappa a)
\sum_{n=0}^\infty(-\lambda)^n I_m^n(\kappa a)K_m^n(\kappa a)\nn\\
&&\times\Bigg\{\left[-\kappa^2+(1-4\xi)\left(\kappa^2+\frac{m^2}r\right)
\right]K_m^2(\kappa r)+(1-4\xi)\kappa^2 K_m^{\prime2}(\kappa r)\Bigg\}.
\eea
If we again use the leading uniform asymptotic expansions for the Bessel
functions we obtain the expression for the leading behavior of the term 
of order $\lambda^{n}$,
\be
u^{(n)}(r)\sim \frac1{8\pi^2r^4}\left(-\frac\lambda2\right)^{n}
\int_0^\infty dz\,z
\sum_{m=1}^\infty m^{3-n}e^{-m\chi}t^{n-1}(t^2+1-8\xi).
\ee
The sum on $m$ is asymptotic to
\be
\sum_{m=1}^\infty m^{3-n}e^{-m\chi}\sim (3-n)!\left(\frac{t r}{2(r-a)}
\right)^{4-n},\quad r\to a+,\label{kam:asymm}
\ee
so the most singular behavior of the order $\lambda^n$ term is, as $r\to a+$,
\be
u^{(n)}(r)\sim (-\lambda)^n\frac{(3-n)!\,(1-6\xi)}{96\pi^2 r^n(r-a)^{4-n}}.
\ee
This is exactly the result found for the weak-coupling limit for a 
$\delta$-sphere\cite{Cavero-Pelaez:2005kq} and for a 
$\delta$-plane,\cite{Milton:2004vy}
so this is a universal result, without physical significance.  It may be
made to vanish by choosing the conformal value $\xi=1/6$.

\subsection{Conformal weak coupling}
With this conformal choice, once again we must expand to higher order.
Besides the corrections noted in \sref{kam:secconf}, we also need
\be
\tilde t\equiv t(z a/r)\sim t+(t-t^3)\frac{r-a}r,\qquad r\to a,
\ee
Then a quite simple calculation gives
\be
u^{(n)}\sim(-\lambda)^n\frac{(n-1)(n+2)\Gamma(3-n)}{2880\pi^2 r^{n+1}
(r-a)^{3-n}},\quad r\to a+,
\ee
which is analytically continued from the region $1\le \mbox{Re}\,n<3$. 
 Remarkably, this
is exactly one-half the result found in the same weak-coupling expansion
for the leading conformal divergence outside a 
sphere.\cite{Cavero-Pelaez:2005kq}
Therefore, like the strong-coupling result, 
this limit is universal, depending on the sum of the principal curvatures 
of the interface.  Note this vanishes for $n=1$, so in every case this
divergence is integrable.
\section{Cylindrical Shell of Finite Thickness}

We now regard the shell (annulus) to have a finite
thickness $\delta$.  We consider the potential
\be
\mathcal{L}_{\rm int}=-\frac\lambda{2a}\phi^2\sigma(r),
\ee
where
\be
\sigma(r)=\left\{\begin{array}{cc}
0,&r<a_-,\\
h,&a_-<r<a_+,\\
0,&a_+<r.\end{array}\right.
\ee
Here $a_\pm=a\pm\delta/2$, and we set $h\delta=1$.  In the limit as
$\delta\to 0$ we recover the $\delta$-function potential.
As for the sphere\cite{Cavero-Pelaez:2005kq} it is straightforward to find the
Green's function for this potential.  In fact, the result may be obtained
from the reduced Green's function given in Ref.~\refcite{Cavero-Pelaez:2005kq} 
by an evident substitution.
Here, we content ourselves by stating the result for the Green's function
in the region of the annulus, $a_-<r,r'<a_+$:
\bea
g_m(r,r')&=&I_m(\kappa'r_<)K_m(\kappa'r_>)+AI_m(\kappa'r)I_m(\kappa'r')\nn\\
&&\mbox{}+B[I_m(\kappa'r)K_m(\kappa'r')+K_m(\kappa'r)I_m(\kappa'r')]
+C K_m(\kappa'r)K_m(\kappa'r'),\nn\\
\eea
where $\kappa'=\sqrt{\kappa^2+\lambda h/a}$.  
The coefficients appearing here are
\begin{subequations}
\bea
A&=&-\frac1\Xi[\kappa I_m'(\kappa a_-)K_m(\kappa' a_-)-\kappa'
I_m(\kappa a_-)K'_m(\kappa'a_-)]\nn\\
&&\quad\times[\kappa K_m'(\kappa a_+)K_m(\kappa' a_+)-\kappa'
K_m(\kappa a_+)K'_m(\kappa'a_+)],\\
B&=&\frac1\Xi[\kappa I_m'(\kappa a_-)I_m(\kappa' a_-)-\kappa'
I_m(\kappa a_-)I'_m(\kappa'a_-)]\nn\\
&&\quad\times[\kappa K_m'(\kappa a_+)K_m(\kappa' a_+)-\kappa'
K_m(\kappa a_+)K'_m(\kappa'a_+)],\\
C&=&-\frac1\Xi[\kappa I_m'(\kappa a_-)I_m(\kappa' a_-)-\kappa'
I_m(\kappa a_-)I'_m(\kappa'a_-)]\nn\\
&&\quad\times[\kappa K_m'(\kappa a_+)I_m(\kappa' a_+)-\kappa'
K_m(\kappa a_+)I'_m(\kappa'a_+)],
\eea
\end{subequations}
where the denominator is
\bea\Xi&=&[\kappa I_m'(\kappa a_-)K_m(\kappa' a_-)-\kappa'
I_m(\kappa a_-)K'_m(\kappa'a_-)]\nn\\
&&\quad\times[\kappa K_m'(\kappa a_+)I_m(\kappa' a_+)-\kappa'
K_m(\kappa a_+)I'_m(\kappa'a_+)]\nn\\
&&\quad\mbox{}-[\kappa I_m'(\kappa a_-)I_m(\kappa' a_-)-\kappa'
I_m(\kappa a_-)I'_m(\kappa'a_-)]\nn\\
&&\quad\times[\kappa K_m'(\kappa a_+)K_m(\kappa' a_+)-\kappa'
K_m(\kappa a_+)K'_m(\kappa'a_+)].
\eea
\subsection{Energy within the shell}
The general expression for the energy
density within the shell is given in terms of these coefficients by
\bea
u(r)&=&\frac1{8\pi^2}\int_0^\infty d\kappa\,\kappa\left[-\kappa^2+(1-4\xi)
\frac1r\frac\partial{\partial r}r\frac\partial{\partial r}\right]\nn\\
&&\quad\times\sum_{m=-\infty}^\infty
[A I_m^2(\kappa' r)+C K_m^2(\kappa'r)+2B K_m(\kappa'r)
I_m(\kappa'r)].\label{kam:enshell}
\eea
\subsection{Leading surface divergence}
The above expressions are somewhat formidable.  Therefore, to isolate
the most divergent structure, we replace the Bessel functions by the leading
uniform asymptotic behavior (\ref{kam:uae}).  A simple calculation implies
\begin{subequations}
\bea
A&\sim&\frac{t_+-t_+'}{t_++t_+'}e^{-2m\eta'_+},\\
B&\sim&\frac{t_+-t_+'}{t_++t_+'}\frac{t_--t_-'}{t_-+t_-'}e^{2m(\eta'_--
\eta'_+)},\\
C&\sim&\frac{t_--t_-'}{t_-+t_-'}e^{2m\eta'_-},
\eea
\end{subequations}
where $t_+=t(z_+)$, $\eta'_-=\eta(z'_-)$, $z_-'=\kappa'a_-/m$, etc.  
If we now insert this approximation into the
form for the energy density, we find
\bea
u&=&\langle T^{00}\rangle=\frac1{8\pi^2a_+^4}2\sum_{m=1}^\infty m\int_0^\infty
dz_+\,z_+ t_r'\nn\\
&&\quad\times\bigg\{\left[\frac{t_+-t_+'}{t_++t_+'}e^{2m(\eta_r'-\eta_+')}
+\frac{t_--t_-'}{t_-+t_-'}e^{2m(-\eta'_r+\eta'_-)}\right]\nn\\
&&\qquad\times\left[\frac{m^2z_+^2}2(1-8\xi)+\left(
\frac{\lambda ha_+^2}a+\frac{m^2a_+^2}{r^2}
\right)(1-4\xi)\right]\nn\\
&&\quad\mbox{}-m^2z_+^2\frac{t_+-t_+'}{t_++t_+'}\frac{t_--t_-'}{t_-+t_-'}
e^{2m(\eta'_--\eta'_+)}\bigg\}.
\eea

If we are interested in the surface divergence as $r$ approaches the outer
radius $a_+$ from within the annulus,
the dominant term comes from the first exponential factor
only.  Because we are considering the limit $\lambda h a\ll m^2$, we
have
\be
t_+'\approx t_+\left(1-\frac{\lambda h}{2m^2}\frac{a_+^2}a t_+^2\right),
\ee
 and we have
\be
u\sim-\frac{\lambda h /a}{32 \pi^2 a_+^2}\sum_{m=1}^\infty m\int_0^\infty
dz\,z t(1-8\xi +t^2)e^{2m(\eta_r'-\eta_+')}.
\ee
The sum over $m$ is carried out according to \eref{kam:asymm}, or
\be
\sum_{m=1}^\infty m e^{2m(\eta'_r-\eta_+')}\sim\left(\frac{r t_r'}{2(r-a_+)}
\right)^2,
\ee
and the remaining integrals over $z$ are elementary.  The result
is
\be
u\sim \frac{\lambda h}{96\pi^2a}\frac{1-6\xi}{(r-a_+)^2},
\quad r\to a_+,
\ee
the expected universal divergence of a scalar field near a
surface of discontinuity,\cite{Bordag:1996zb} without significance,
which may be eliminated by setting $\xi=1/6$.

\subsection{Surface energy}

Now we want to establish that the surface energy $\mathfrak{E}$
(\ref{kam:surfenergy}) is the same as the integrated local energy density in 
the annulus when the limit $\delta\to 0$ is taken.  To examine this limit,
we consider $\lambda h/a\gg \kappa^2$.  So we apply the uniform
asymptotic expansion for the Bessel functions of $\kappa'$ only. We must
keep the first two terms in powers of $\kappa\ll\kappa'$:
\bea
\Xi&\sim&-\kappa^{\prime2}\frac{I_m(\kappa a_-)K_m(\kappa a_+)}{mz_-'z_+'
\sqrt{t_-'t_+'}}\sinh m(\eta_-'-\eta_+')\nn\\
&&\!\!\!\!\!\!-\frac{\kappa'\kappa}m\left[\frac1{z_+'}\sqrt{\frac{t_-'}{t_+'}}I_m'(\kappa
a_-)K_m(\kappa a_+)-\frac1{z_-'}\sqrt{\frac{t_+'}{t_-'}}I_m(\kappa
a_-)K_m'(\kappa a_+)\right]\nn\\
&&\qquad\times\cosh m(\eta'_--\eta_+').
\eea
Because we are now regarding the shell as very thin,
\be
\eta'_--\eta'_+\approx -\frac\delta a\frac1{t'},
\ee
where
\be
t'\sim \frac1{z'}\sim \frac{m}{\sqrt{\lambda h a}},
\ee
 using the Wronskian (\ref{kam:wronskian}) we get the denominator
\be
\Xi\sim -\frac1{a^2}[1+\lambda I_m(\kappa a)K_m(\kappa a)].
\ee
Then we immediately find the interior coefficients:
\begin{subequations}
\bea
A&\sim&\frac\pi2\sqrt{\lambda h a}\frac{I_m(\kappa a)K_m(\kappa a)}{1+\lambda
I_m(\kappa a)K_m(\kappa a)}e^{-2m\eta'},\\
B&\sim&\frac12\sqrt{\lambda h a}\frac{I_m(\kappa a)K_m(\kappa a)}{1+\lambda
I_m(\kappa a)K_m(\kappa a)},\\
C&\sim&\frac1{2\pi}\sqrt{\lambda h a}\frac{I_m(\kappa a)K_m(\kappa a)}{1+\lambda
I_m(\kappa a)K_m(\kappa a)}e^{2m\eta'}.
\eea

\end{subequations}
\subsection{Identity of shell energy and surface energy}
We now insert this in the expression for the energy density (\ref{kam:enshell})
 and keep only the largest terms, thereby neglecting $\kappa^2$ relative to
$\lambda h/a$.  This gives a leading term proportional to $h$, which
when multiplied by the area of the annulus $2\pi a\delta$ gives for
the energy in the shell
\be
\mathcal{E}_{\rm ann}\sim(1-4\xi)\frac{\lambda}{4\pi a^2}
\sum_{m=-\infty}^\infty \int_0^\infty d\kappa a\,\kappa a
\frac{I_m(\kappa a)K_m(\kappa a)}{1+\lambda I_m(\kappa a)K_m(\kappa a)},
\ee
which is exactly the form of the surface energy $\mathfrak{E}$
given by the negative of the second term in the integrated energy density
(\ref{kam:inten}).
\subsection{Renormalizability of surface energy}
In particular, note that the term in $\mathfrak{E}$ of order $\lambda^3$
is, for the conformal value $\xi=1/6$, exactly equal to that term in the
total energy $\mathcal{E}$ in \eref{kam:thirddiv}:
\be
\mathfrak{E}^{(3)}=\mathcal{E}^{(3)}.
\ee
This means that the divergence encountered in the global energy
is exactly accounted for by the divergence in the surface energy, which
would seem to provide strong evidence in favor of the renormalizablity
of that divergence.  

\section{Conclusion}

The work reported here and in 
Refs.~\refcite{Cavero-Pelaez:2005kq,Cavero-Pelaez:2006rt} represents
a significant advance in understanding the divergence structure of
Casimir self-energies.  We have shown that the surface energy of a
$\delta$-function shell potential is in fact the integrated local energy
density contained within the shell when the shell is given a finite
thickness.  That surface energy contains the entire third-order divergence
in the total Casimir energy.  The local Casimir energy diverges as the
shell is approached, but that divergence is integrable, so it yields a
finite contribution to the total energy.  The identification of the
divergent part of the total energy with that associated with the surface
strongly suggests that this divergence can be absorbed in a renormalization
of parameters describing the background potential.

Challenges yet remain.  This renormalization procedure needs to be made
precise.  Further, we must make more progress in understanding the sign
(and for cylindrical geometries, the vanishing) of the total Casimir 
self-energy. And, of course, we must understand the implications of surface
divergences on the coupling to gravity. Work is proceeding in all
these directions.

\section*{Acknowledgments}
We thank the US National Science Foundation and the US Department of Energy
for partial funding of this research.  KAM is grateful to Vladimir Mostepanenko
for inviting him to participate in MG11.  We thank S. Fulling, P. Parashar, 
A. Romeo, K. Shajesh, and J. Wagner for useful discussions.

%\bibliographystyle{ws-procs975x65}
%\bibliography{casimir}

\end{document}